\documentclass[showpacs,amsmath,amssymb,prd,floatfix,twocolumn]{revtex4}

\usepackage{epsfig}
\usepackage{graphicx}
\usepackage{bm}
\usepackage{amsfonts}

\def\bfig{\begin{figure}[ht] \begin{center}}
\def\bfigh{\begin{figure}[h!] \begin{center}}
\def\bfigb{\begin{figure}[hb!] \begin{center}}
\def\bfigt{\begin{figure}[t!] \begin{center}}
\def\bfight{\begin{figure}[ht!] \begin{center}}
\def\efig{\end{center} \end{figure}}

\def\btab{\begin{table*}[ht]}
\def\etab{\end{table*}}

\begin{document}

\title{Phantom evolution in power-law potentials}

\author{Emmanuel N.~Saridakis }
\email{msaridak@phys.uoa.gr} \affiliation{Department of Physics,
University of Athens, GR-15771 Athens, Greece}

\begin{abstract}
We investigate phantom models with power-law potentials and we
extract the early-time, ``tracker'', solutions under the
assumption of matter domination. Contrary to quintessence case, we
find that energy positivity requires normal power-law potentials
instead of inverse power-law ones,  with the potential exponent
being bounded by the quadratic form. In addition, we analytically
present the general cosmological solution at intermediate times,
that is at low redshifts, which is the period of the transition
from matter to dark-energy domination. The comparison with the
exact evolution, arising from numerical elaboration, shows that
the tracker solution agrees with the later within $2\%$ for
redshifts $z\gtrsim1.5$, while the intermediate solution is
accurate within $2\%$ up to $z\approx0.5$.
\end{abstract}

 \maketitle

\section{Introduction}

Recent cosmological observations support that the universe is
experiencing an accelerated expansion, and that the transition to
the accelerated phase has been realized in the recent cosmological
past \cite{observ}. In order to explain this remarkable behavior,
and despite the intuition that this can be achieved only through a
fundamental theory of nature, physicists can still propose some
paradigms for its description, such are theories of modified
gravity \cite{ordishov}, or ``field'' models of dark energy.  The
field models that have been discussed widely in the literature
consider a canonical scalar field (quintessence)
\cite{quint0,quint}, a phantom field, that is a scalar field with
a negative sign of the kinetic term \cite{phant}, or the
combination of quintessence and phantom in a unified model named
quintom \cite{quintom}.

In field dark energy models one can find potential-independent
solutions \cite{Scherrer:2007pu}, but in general  the cosmological
evolution depends significantly on the potential choice. In the
quintessence case, one well studied potential is the inverse-power
law one \cite{quint0,Watson:2003kk,Kneller:2003xg}. This potential
exhibits ``tracker'' behavior, that is for a large class of
initial conditions the cosmological evolution converges to a
common solution at late times \cite{quint01}. Furthermore, in such
models the quintessence energy density remains small at early and
intermediate times and thus  the known cosmological epochs are not
affected. Finally, a theoretical justification of power-law
potentials can arise through supersymmetric considerations
\cite{Binetruy:1998rz}.

The cosmological evolution in such quintessence models at early
times, is dominated by the matter-fluid and one can obtain the
tracker solutions analytically \cite{quint01}. At intermediate
times the quintessence field is still small, but non-negligible,
and one has to rely on perturbative analytical expressions
\cite{Watson:2003kk}. However, such an extension to low redshift
is very useful, since observations like supernovae Ia, WMAP and
SDSS ones, are related to this period \cite{observ}.

In this work we are interested in investigating the phantom
scenario in power-law potentials, both at high and low redshifts,
and provide the tracker solutions and the perturbative analytical
expressions respectively. Due to the ``inverse'' kinetic behavior
of phantoms in potential slopes, we expect that we have to
consider normal power-law  potentials instead of inverse power-law
ones. Indeed, starting with an arbitrary power law, we find that
physically meaningful cosmological evolution excludes negative
exponents.

The plan of the work is as follows: In section \ref{zeroord} we
construct phantom models in power-law potentials and we present
the matter-dominated solutions. In section \ref{firstord} we
derive the cosmological solutions at intermediate times, that is
when the phantom dark energy is non-negligible, but still
sub-dominant comparing to the matter content of the universe. In
section \ref{analnum} we compare our analytical expressions with
the exact numerical evolution of the phantom scenario. Finally, in
section \ref{discuss} we provide some applications of the model
discussing its cosmological consequences, and we summarize our
results.

\section{Cosmological evolution during matter domination} \label{zeroord}

Let us construct the simple phantom cosmological scenario in a
flat universe. The action of a universe constituted of a phantom
field $\phi$ is \cite{phant}:
\begin{eqnarray}
S=\int d^{4}x \sqrt{-g} \left[\frac{1}{2} R
+\frac{1}{2}g^{\mu\nu}\partial_{\mu}\phi\partial_{\nu}\phi+V(\phi)
+\cal{L}_\text{m}\right], \label{actionquint}
\end{eqnarray}
where we have set $8\pi G=1$.  $V(\phi)$ is the phantom field
potential and the term $\cal{L}_\text{m}$ accounts for the (dark)
matter content of the universe, considered as dust. Finally,
although we could straightforwardly include baryonic matter and
radiation in the model, for simplicity reasons we neglect them
since we are interested in $z<20$ era. The Friedmann equations
are \cite{phant}:
\begin{equation}\label{FR1}
H^{2}=\frac{1}{3}\Big(\rho_{m}+\rho_{\phi}\Big),
\end{equation}
\begin{equation}\label{FR2}
\dot{H}=-\frac{1}{2}\Big(\rho_{m}+\rho_{\phi}+p_{\phi}\Big),
\end{equation}
while the evolution equation for the phantom field is:
\begin{equation}
\label{eom} \dot{\rho}_\phi+3H(\rho_\phi+p_\phi)=0,
\end{equation}
where $H=\dot{a}/a$ is the Hubble parameter and $a = a(t)$ the
scale factor. In these expressions, $\rho_m$ is the dark matter
density, and  $\rho_{\phi}$ and $p_\phi$  are respectively the
density and pressure of the phantom field given by:
\begin{eqnarray}\label{rhosigma}
 \rho_{\phi}&=& -\frac{1}{2}\dot{\phi}^{2} + V(\phi)\\
 p_{\phi}&=& - \frac{1}{2}\dot{\phi}^{2} - V(\phi).\label{psigma}
\end{eqnarray}
 Therefore,
we can equivalently write the  evolution equation in field terms
as:
\begin{equation}
\label{phiddot}
 \ddot{\phi}+3H\dot{\phi}-\frac{d
V(\phi)}{d\phi}=0.
\end{equation}
The system of equations closes by considering the evolution of the
matter density, which in the case of dust reads simply
$\rho_m=\rho_{m0}/a^3$, with $\rho_{m0}$ its value at present.
Finally, the dark energy equation-of-state parameter is given by
\begin{equation}
w\equiv\frac{p_\phi}{\rho_\phi}.
\end{equation}

As mentioned in the introduction, in the present work we study
phantom evolution in power-law potentials. That is we consider:
\begin{equation}
V(\phi) = \kappa\phi^{-\alpha},
 \label{pot}
\end{equation}
where $\kappa$ is a constant with units of $m^{4+\alpha}$. Note
that we prefer to insert the minus sign in the exponent in order
to coincide with the usual notation of the literature. A priori we
do not restrict the value of $\alpha$. However, as we are going to
see, energy positivity will force $\alpha$ to be negative and
bounded, that is potential (\ref{pot}) will be a normal (and not
inverse) power law.

It proves convenient to express the aforementioned cosmological
system using the scale factor $a$ as the independent variable,
since it is straightforwardly related to the redshift $z$ which is
used in observations. Following \cite{Watson:2003kk} we define
\begin{equation}
 x =
{\rho_\phi + p_\phi\over 2(\rho_m + \rho_\phi)} = {-{1\over 2}
\dot\phi^{2}\over 3H^2} = -{1\over 6}{\Big(a{d\phi\over
da}\Big)}^2, \label{xdef}
\end{equation}
and thus the first Friedmann equation gives
 \begin{equation}
-\frac{1}{2} \dot\phi^2 = {x\over 1-x}(\rho_m + V),
\label{phidotx}
\end{equation} where $V$ stands for $V(\phi)$. Thus, one can
simply write:
\begin{eqnarray}
 &&\rho_{\phi} = \frac{x\rho_m + V}{1-x}\label{rhox}\\
 &&p_{\phi}=\frac{x\rho_m - V(1-2x)}{1-x}\label{px}\\
&&3H^2 = \frac{\rho_m + V}{1-x}\label{Hx}\\
 && w = \frac{x\rho_m - V(1-2x)}{x\rho_m + V}\label{wx}.
\end{eqnarray}
Comparing to the quintessence case, and apart from the obvious
expectation that $x$ is now negative, we observe, that it is also
bounded. In particular, requiring phantom total energy positivity
($\rho_\phi>0$) we obtain $x>-V/(3H^2)$. Finally, using
expressions (\ref{rhox})-(\ref{Hx}), the field evolution equation
(\ref{phiddot}) becomes
\begin{equation}
a^{2}{\phi}'' + {a{\phi}'\over 2}\left(5-3x\right) + {3(1-x)\over
\rho_m +V}\Big({a{\phi}'V\over 2} - {dV\over d\phi}\Big) = 0,
\label{eomx}
\end{equation}
with prime denoting the derivative with respect to $a$. This
equation is exact and accounts for the complete  dynamics of the
phantom scenario.

In this section we desire to present the tracker solutions during
the matter-dominated era. Thus, we consider
$\rho_\phi,|p_\phi|\ll\rho_m$, or equivalently $|x|\ll1$. In this
case equation (\ref{eomx}) is simplified as:
\begin{equation}
a^{2}{\phi}''_{(0)} + {5a{\phi}'_{(0)}\over 2} - {3\over
\rho_m}{dV\over d\phi} = 0, \label{eomxzero}
 \end{equation}
with $\frac{dV}{d\phi}\equiv
\frac{dV}{d\phi}\Big|_{\phi=\phi_{(0)}}$. The zero subscript in
parentheses denotes just this zeroth-order solution in terms of
$x$, or equivalently in terms of the phantom energy density, and
must not be confused later on with the subscript $0$ without
parentheses which stands for the present value of a quantity.

Equation (\ref{eomxzero}) can be easily solved analytically in the
case of power-law potentials. For $\alpha=-2$ we acquire:
\begin{equation}
\phi_{(0)} =
e^{-2\sqrt{\frac{2\kappa}{3\rho_{m0}}}a^{3/2}}\Big(c_1\,e^{4\sqrt{\frac{2\kappa}{3\rho_{m0}}}a^{3/2}}-c_2\Big)a^{-3/2},
\label{solalpha2}
\end{equation}
where $c_1$ and $c_2$ are constants determined by initial
conditions. For $\alpha\neq-2$ and $\alpha\neq0$ (which is the
trivial case of a constant potential), the general solution of
(\ref{eomxzero}) is
\begin{equation}
 \phi_{(0)} = C(\alpha){a}^{{3/(2 + \alpha)}},
\label{solphizero}
\end{equation}
where the function $C(\alpha)$ is related to the potential
parameters through
\begin{equation}
C(\alpha) =\Big[-\frac{2\alpha
(2+\alpha)^2\kappa}{3\rho_{m0}(4+\alpha)}\Big]^{\frac{1}{2+\alpha}}.
 \label{Calpha}
\end{equation}
In (\ref{solphizero}) we have kept only the solution part that
remains small (together with its derivative) for small $a$'s, in
order to be consistent with the matter-dominated  approximation
($\rho_\phi,|p_\phi|\ll\rho_m$). That is why initial-condition
dependent constants are absent (which is just the central idea of
tracker solutions). In addition, it is easy to see that this
solution remains regular for $\alpha\rightarrow-2$.

 The zeroth-order solution for
$\rho_\phi$ can be calculated from (\ref{rhox}) under $x\ll1$,
that is
 $\rho_{\phi (0)} =x_{(0)}\rho_m+V( \phi_{(0)})$, where $x_{(0)} =-
(a\phi'_{(0)})^{2}/6$. The result is:
 \begin{equation}
\label{rhophizero}
 \rho_{\phi (0)} =-
3\rho_{m0}{{C(\alpha)^2}\over {\alpha(2 +
\alpha)}}\,a^{-\frac{3\alpha}{2 + \alpha}}.
\end{equation}
From this expression we educe the physically interesting result
that $-2<\alpha<0$ in order for $ \rho_{\phi (0)}$ to be positive
(additionally in this case the $C(\alpha)$-definition in
(\ref{Calpha}) is unambiguous). The restriction of $\alpha$ to
negative values, that is the consideration of normal power-law
potentials instead of inverse power-law ones (contrary to
quintessence case), was expected due to the ``inverse'' kinetic
behavior of phantoms in potential slopes. The additional lower
bound of $\alpha$ values, which breaks this form of symmetry
between phantom and quintessence models, arises from the extra and
necessary assumption of positivity of the total phantom energy,
which is not present in quintessence case
\cite{quint0,Watson:2003kk,Kneller:2003xg} where one can always
obtain it (the potentials can always be shifted to positive values
in cosmology). In other words, if the potential is too steep
(large $-\alpha$), then at early times (where the field value
$\phi\ll1$, i.e it is close to the potential minimum) the phantom
potential energy will be very small ($\propto \phi^{-\alpha}$) and
thus unable to make the total energy positive. On the other hand,
if the potential is smooth (small $-\alpha$) then positivity is
obtained. Quantitatively, $\alpha=-2$ separates the two regions,
and indeed one can see that the energy density arising form
(\ref{solalpha2}) is negative unless $c_1,c_2\rightarrow0$, in
which case it is exactly zero. But $c_1,c_2\rightarrow0$ is
implied by the requirement $\phi_{(0)},\phi_{(0)}'\ll1$ at small
$a$. Therefore, for the limiting case $\alpha=-2$, the
matter-dominating condition leads to $\rho_{\phi(0)}=0$, and thus
this exponent value bounds the two aforementioned regions.

 The corresponding zeroth-order solution for
$p_\phi$ is
 $p_{\phi (0)} =x_{(0)}\rho_m-V( \phi_{(0)})$, which leads to:
 \begin{equation}
\label{pphizero}
 p_{\phi (0)} =
6\rho_{m0}{{C(\alpha)^2}\over {\alpha(2 +
\alpha)^2}}\,a^{-\frac{3\alpha}{2 + \alpha}}.
\end{equation}
Therefore, for the zeroth-order solution of the equation-of-state
parameter we obtain:
\begin{equation}
\label{wzerosol}
   w_{(0)} =\frac{ p_{\phi (0)}}{\rho_{\phi (0)}}= -{2\over
   2+\alpha},
 \end{equation}
 that is $w_{(0)}$ is constant. Note that $-2<\alpha<0$ leads to $ w_{(0)}<-1$, which is a
self-consistency test of our calculations since it is the basic
feature of phantom cosmology. Finally, we can calculate the
 zeroth-order behavior of the phantom density
parameter as $\Omega_{\phi (0)} = \rho_{\phi (0)}/(\rho_{\phi (0)}
+ \rho_{m})$, with $\rho_m=\rho_{m0}/a^3$, acquiring
\begin{equation}
 \Omega_{\phi(0)}=1-\left[1-\frac{ 3C(\alpha)^2 }{\alpha(2+\alpha)} a^{\frac{6}{2+\alpha}}\right]^{-1}\label{Om0tilde}.
\end{equation}

Expressions (\ref{solphizero}), (\ref{rhophizero}),
(\ref{wzerosol}) and (\ref{Om0tilde}) are the tracker solutions
for phantom cosmology in power-law potentials. Equivalently they
can be expressed as a function of time, considering $a\propto
t^{2/3}$
 since we are in the matter-dominated era, as:
 \begin{eqnarray}
\label{trackersol1}
 &&\phi \propto t^{2/(2+\alpha)}\\
 &&\label{trackersol2}
\rho_\phi \propto t^{-2 \alpha/(2+\alpha)},
\end{eqnarray}
with the equation-of-state parameter given by (\ref{wzerosol}).
These expressions provide an excellent approximation to the
behavior of the phantom field as long as $\rho_\phi,|p_\phi|
\ll\rho_m$.

\section{ First-order perturbation of matter-dominated  cosmological evolution} \label{firstord}

As cosmological evolution continues, the phantom field increases
and phantom energy density becomes non-negligible, although still
dominated by the dark matter one. Thus, we expect that the various
quantities will progressively start diverging from the expressions
obtained above. In this section we are interested in studying
analytically this intermediate evolution stage, and thus we
perform a first-order perturbation to the zeroth order solutions
of section \ref{zeroord}. The corresponding  quantities are
denoting by the subscript $1$ in parentheses, and the total ones
by tilde, i.e:
\begin{equation}
\label{firstorder}
 \widetilde\phi = \phi_{(0)} + \phi_{(1)}, \  \widetilde\rho_\phi = \rho_{\phi{(0)}} +
 \rho_{\phi{(1)}},\  \widetilde w = w_{(0)} + w_{(1)}.
 \end{equation}
This perturbation is equivalent with keeping terms up to first
order in $x$ (i.e keeping only $x_{(0)}$) and in
$\phi_{(1)}/\phi_{(0)}$, in the exact evolution equation
(\ref{eomx}). In addition, one can also expand the potential as
$V(\widetilde\phi)=V(\phi_{(0)})+\frac{d V(\phi_{(0)})}{d
\phi_{(0)}} \phi_{(1)}+{\mathcal{O}}( \phi_{(1)}^2)$. Therefore,
inserting the expansions (\ref{firstorder}) in (\ref{eomx}),  we
obtain
\begin{eqnarray}
 a^{2}{\phi_{(1)}''} + {a\over
2}\left[5\phi_{(1)}'-3x_{(0)}\phi_{(0)}'\right] +
{3a\phi_{(0)}'V\over 2 \rho_{m}}+
\nonumber \\
+{3\rho_{\phi (0)}\over  {(\rho_{m})}^{2}} {dV\over d\phi}- {3
\over \rho_{m}}{d^2 V \over d\phi^2}\phi_{(1)} = 0,
\label{eomxfristord}
 \end{eqnarray}
  where $x_{(0)} =-
(a\phi'_{(0)})^{2}/6$. Thus, for the power-law potential
(\ref{pot}) we acquire
 \begin{eqnarray}
a^{2}{\phi_{(1)}''} &+& {a\over
2}\left[5\phi_{(1)}'+\frac{1}{2}a^2(\phi_{(0)}')^3\right] + {27
\kappa a^{3} \phi_{(0)}^{1-\alpha}\over 2(2+\alpha)\rho_{m0}}-
\nonumber \\
&-& {3\alpha(1+\alpha)\kappa a^{3}
\phi_{(0)}^{-2-\alpha}\phi_{(1)}\over \rho_{m0}} = 0.
\label{eomxfristord2}
\end{eqnarray}
 The general solution of (\ref{eomxfristord2}), for $-2<\alpha<0$,
 is \begin{equation} \phi_{(1)} =  {3(6+\alpha)\over
\alpha(2+\alpha)(\alpha^2+8\alpha+28)}\,\phi_{(0)}^{3},
\label{phifirstorder}
\end{equation}
where we have kept only the  part that remains small (together
with its derivative) for small $a$'s, in order to be consistent
with the matter-dominated  approximation.

The  perturbation of $\rho_\phi$ can be calculated from
(\ref{rhox}) keeping the corresponding terms, thus:
\begin{equation}
\rho_{\phi (1)} = x_{(0)}\rho_{\phi (0)} + x_{(1)}\rho_m -
\alpha{\phi_{(1)}\over \phi_{(0)}}V,\label{rhophifirstorder}
\end{equation}
where $x_{(1)} = -\frac{1}{3}a^2\phi_{(0)}'\phi_{(1)}'$ as it
arises from $\widetilde{x} = x_{(0)} + x_{(1)}=-
a^{2}(\phi'^{2}_{(0)} + 2\phi'_{(0)}\phi'_{(1)})/6$. Therefore,
using also (\ref{solphizero}), (\ref{rhophizero}) and
(\ref{phifirstorder}) in order to express the result in terms of
$\phi_{(0)}$ and $\phi_{(1)}$, we acquire
\begin{equation}
 \rho_{\phi (1)} = -\frac{\alpha(4+\alpha)}
{6+\alpha}\Big( {\phi_{(1)}\over \phi_{(0)}}\Big)\rho_{\phi (0)}.
\label{rhophifirstordersolna}
 \end{equation}
Similarly, the  perturbation of $w$ can arise form (\ref{wx}) as
\begin{equation}
w_{(1)} =\frac{p_{\phi (1)}}{\rho_{\phi (0)}}-\frac{p_{\phi
(0)}}{\rho_{\phi (0)}^2} \rho_{\phi (1)}=
 (1-w_{(0)})\Big({\rho_{\phi (1)}\over \rho_{\phi
(0)}} + \alpha{\phi_{(1)}\over \phi_{(0)}}\Big),
\label{wfirstorder}
\end{equation}
since $p_{\phi (1)} = x_{(0)}\rho_{\phi (0)} + x_{(1)}\rho_m +
\alpha{\phi_{(1)}\over \phi_{(0)}}V$. Therefore, the final result
writes
 \begin{equation}
  w_{(1)} = -\frac{\alpha(4+\alpha)}
{6+\alpha}\Big({\phi_{(1)}\over \phi_{(0)}}\Big)w_{(0)}.
\label{wfirstordersolna}
\end{equation}
Expressions (\ref{rhophifirstordersolna}) and (\ref{wfirstorder})
coincide with the corresponding one for quintessence in inverse
power-law potentials \cite{Watson:2003kk}, however
${\phi_{(1)}\over \phi_{(0)}}$ and $\rho_{\phi (0)}$ has a sign
difference and in addition $\alpha$ is negative and bounded. Thus,
$\phi_{(1)}$ is negative and, contrary to quintessence,
$\rho_{\phi (1)}$ is negative while $w_{(1)}$ is positive.

In conclusion, the total solutions for intermediate times, that is
at low redshifts, read:
\begin{eqnarray}
\widetilde \phi &=&
\phi_{(0)} +  \phi_{(1)}=\nonumber\\
& =& \phi_{(0)}\Big[1 + {3 (6 + \alpha)C(\alpha)^2\over \alpha (2
+ \alpha) (\alpha^2+8\alpha+28)} a^{6/(2+\alpha)}\Big],\ \
\label{phitilde}
\end{eqnarray}
\begin{eqnarray}
\widetilde{\rho}_\phi &=&
\rho_{\phi(0)} + \rho_{\phi(1)}=\nonumber\\
\label{rhotilde} &=& \rho_{\phi (0)}\Big[1 - {
3(4+\alpha)C(\alpha)^2\over (2+\alpha)(\alpha^2+8\alpha+28)}
a^{6/(2+\alpha)}\Big],\ \
\end{eqnarray}
and
\begin{eqnarray}
\widetilde w &=&
w_{(0)} + w_{(1)}=\nonumber \\
\label{wtilde} &=& w_{(0)}\Big[1 -  { 3(4+\alpha)C(\alpha)^2\over
(2+\alpha)(\alpha^2+8\alpha+28)} a^{6/(2+\alpha)}\Big],
\end{eqnarray}
where $C(\alpha)$ is given by (\ref{Calpha}). Finally,
$\widetilde{\Omega}_{\phi}$ can be easily calculated as
 $\widetilde{\Omega}_{\phi} =
\widetilde{\rho}_{\phi}/(\widetilde{\rho}_{\phi} + \rho_{m})$,
giving {\small{
\begin{equation}
\widetilde \Omega_\phi=1-\left\{1+\frac{\rho_{\phi
(0)}}{\rho_{m0}}\left[a^3-{ 3(4+\alpha)C(\alpha)^2\over
(2+\alpha)(\alpha^2+8\alpha+28)}
a^{\frac{3\alpha+12}{2+\alpha}}\right]
 \right\}^{-1}\label{Omtilde}.
\end{equation}}}
Lastly, note that since we have these quantities as a function of
the scale factor, it is straightforward to express them as a
function of the redshift $z$ through $\frac{a_0}{a}=1+z$, with
$a_0=1$ the present value.

\section{Comparing analytical and numerical results}\label{analnum}

In the previous two sections we extracted analytical expressions
for phantom evolution in power-law potentials, when the universe
is dominated by the dark matter sector, and we argued that these
solutions are valid at early and intermediate times, that is at
both high and low redshift. In order to test the precision of our
formulae, we evolve numerically the exact cosmological system
calculating the exact quantities $\phi(a)$, $\Omega_\phi(a)$,
$w(a)$, and then we examine their divergence from the zeroth and
first order expressions by investigating the corresponding ratios.
We choose initial conditions consistent with initial matter
domination (if this requirement is fulfilled then the results do
not depend on the specific initial conditions), and we fix
$\kappa$ in order to acquire $\Omega_{m0} \approx 0.28$ and
$\Omega_{\phi0} \approx 0.72$ at present.

\begin{figure}[ht]
\begin{center}
\mbox{\epsfig{figure=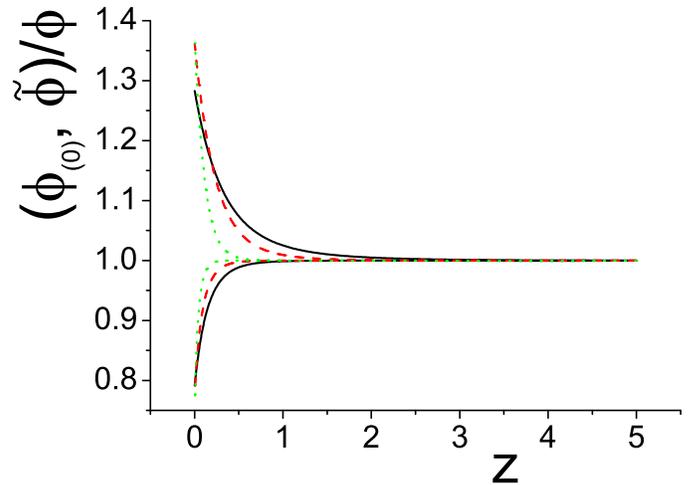,width=9.cm,angle=0}} \caption{
(Color Online){\it The ratios $\phi_{(0)}/\phi$ (upper three
curves) and $\widetilde \phi/\phi$ (lower three curves) as a
function of the redshift, for $\alpha = -0.5$ (black, solid),
$\alpha = -1$ (red, dashed), and $\alpha = -1.5$ (green,
dotted).}} \label{fig1phi}
\end{center}
\end{figure}

In fig. \ref{fig1phi} we present the ratios of the zeroth
(tracker) and  first order  field solutions  to the exact
numerical value, that is $\phi_{(0)}/\phi$ and
$\widetilde{\phi}/\phi$, as a function of the redshift.
 The
calculations have been performed for three potential cases, namely
$\alpha = -0.5$, $\alpha = -1$ and $\alpha = -1.5$. As we observe,
$\phi_{(0)}/\phi$ is very close to $1$  for $z\gtrsim1.5$, that is
the tracker solution is a very good approximation at this early
evolution stage. For $1.5\gtrsim z\gtrsim0.5$ the zeroth order
solution is not a good approximation (with error $8\%$) but the
first order one remains within $98\%$ accuracy (with even better
results for larger $|\alpha|$ values). However, after that stage,
the phantom cosmological effects become significant and our
approximation breaks down rapidly.

\begin{figure}[ht]
\begin{center}
\mbox{\epsfig{figure=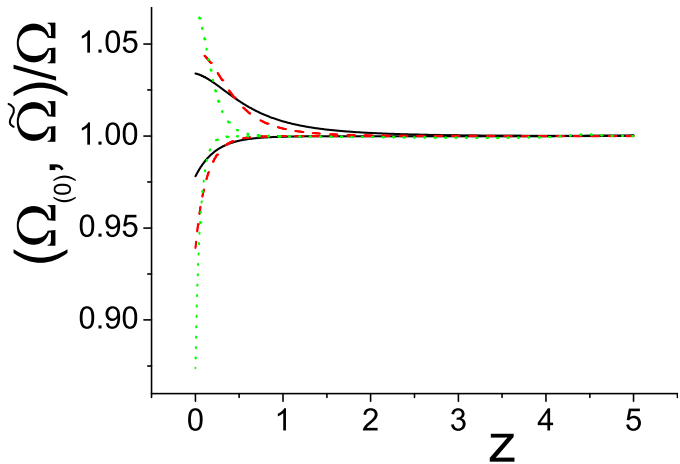,width=9.cm,angle=0}} \caption{
(Color Online){\it The ratios $\Omega_{\phi(0)}/\Omega_\phi$
(upper three curves) and $\widetilde\Omega_{\phi}/\Omega_\phi$
(lower three curves) as a function of the redshift, for $\alpha =
-0.5$ (black, solid), $\alpha = -1$ (red, dashed), and $\alpha =
-1.5$ (green, dotted).}} \label{fig2om}
\end{center}
\end{figure}

Similarly, in fig. \ref{fig2om} we present
$\Omega_{\phi(0)}/\Omega_\phi$  and
$\widetilde\Omega_{\phi}/\Omega_\phi$ versus z.
The divergence of the zeroth order solution from the exact one
starts at $z\approx1.5$. However, as we observe, the first order
solution is very accurate (with error less than $2\%$) up to
$z\approx0.3$. After that point, the first order solution starts
diverging rapidly from the exact one and our approximation is not
valid. The fact that at $z=0$ in some cases (large $|\alpha|$) the
zeroth order solution seems to be closer to the exact evolution
comparing to the first order one, is a result of the signs of the
corresponding terms (namely positive $\phi_{(0)}$ and negative
$\phi_{(1)}$), but it has no meaning since our approximations are
not valid in that regime.

\begin{figure}[ht]
\begin{center}
\mbox{\epsfig{figure=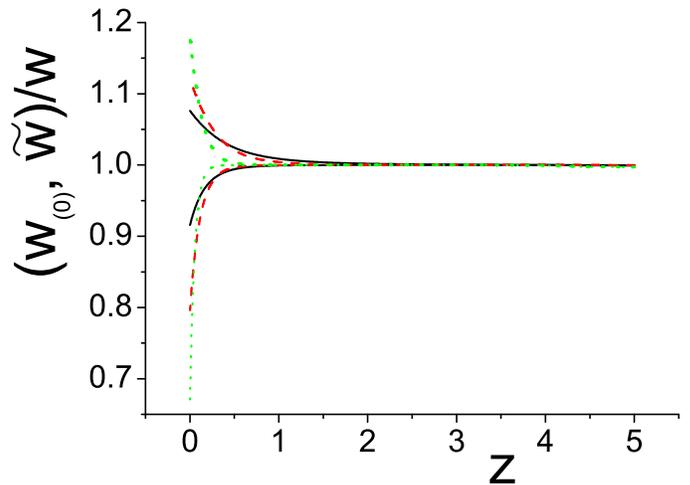,width=9.cm,angle=0}} \caption{
(Color Online){\it The ratios $w_{(0)}/w$ (upper three curves) and
$\widetilde w/w$ (lower three curves) as a function of the
redshift, for $\alpha = -0.5$ (black, solid), $\alpha = -1$ (red,
dashed), and $\alpha = -1.5$ (green, dotted).}} \label{fig3w}
\end{center}
\end{figure}

Finally, in fig. \ref{fig3w}  we depict $w_{(0)}/w$ and
$\widetilde w/w$ as a function of the redshift.
As we see, the tracker solution is very accurate up to
$z\approx1$, while the first order one agrees with the exact
evolution within $2\%$ up to $z\approx 0.3$.

In summary, we observe that the tracker (zeroth order) solution is
accurate within an error  of $2\%$ at early times and up to
$z\approx1.5$. At intermediate times  we have to use the first
order analytical solution, which agrees with the exact
cosmological evolution within an error of $2\%$ up to
$z\approx0.5$. After that point, the phantom dark energy sector
enhances, it dominates the cosmological evolution, and our
approximation breaks down rapidly.

\section{Discussion and Conclusions} \label{discuss}

Since we have tested the accuracy of our analytical expressions
and we have determined their applicability region, one can use
them to describe an arbitrary phantom evolution in power-law
potentials up to $z=0.5$. Here we will present some additional
applications concerning observable quantities.

Having extracted expressions for $w(a)$ and $\Omega_\phi(a)$ which
are valid up to low redshifts, we can derive the corresponding
$w(\Omega_\phi)$ relation . Indeed, using (\ref{rhophizero}),
(\ref{wtilde}) and (\ref{Omtilde}) we result to
\begin{equation}
\label{wOmega} w(\Omega_\phi) = -{2 \over 2+ \alpha} - {2 \alpha
(4 + \alpha) \over (2+\alpha)(\alpha^2+8\alpha+28)} \Omega_\phi.
\end{equation}
Thus, at early times, where phantom energy density (that is
$\Omega_\phi$) is negligible, $w$ is equal to a constant value,
namely $w_{(0)}$. At intermediate times it increases linearly with
$\Omega_\phi$. Comparing with the corresponding result for
quintessence in inverse power-law potentials \cite{Watson:2003kk}
there are two differences. Firstly, $w(\Omega_\phi)$ is an
increasing function instead of a decreasing one. Secondly, since
$\alpha$ lies now in the interval $(-2,0)$, the slope is not a
slowly-varying function of $\alpha$ but it ranges from $0.1$ at
$\alpha=-0.5$ to $0.8$ at $\alpha=-1.5$. Finally, one can examine
the accuracy of expression (\ref{wOmega}) by comparing it to the
exact evolution. Indeed, we find that it is very satisfactory
(with an error of less than $2\%$) up to $z\approx0.5$ (or
equivalently up to $\Omega_\phi\approx0.3$), and then the error
grows reaching to $\approx 9\%$ at $z=0$.

Additionally, and similarly to \cite{Watson:2003kk}, we can use
expression (\ref{wtilde}) for $\widetilde w(a)$ in order to
acquire a parametrization for $w(z)$ for redshifts up to
$z\approx0.5$, where our approximation is still valid. Indeed we
find the same result with \cite{Watson:2003kk}, namely:
\begin{equation}
\label{wfit} w_{\rm fit} = w_{(0)} + (w_0 -
w_{(0)})(1+z)^{-6/(2+\alpha)},
\end{equation}
with the obvious difference that now the undetermined parameter
$w_0$, which is just the present value of $w$, is smaller than
$-1$. This fit for phantom evolution in power-law potentials
proves to be satisfactory within an error of $3\%$. Finally, one
can examine (\ref{wfit}) in parallel with the phenomenological
parametrization $w_{z} = w_0 + w_1 \ln (1+z)$ of
\cite{Efstathiou}, where he finds that the two expressions have a
comparable level of accuracy \cite{Watson:2003kk}.

In this work we have investigated phantom models with power-law
potentials. We have extracted the tracker solutions under the
assumption of matter domination, that is the general solution and
common behavior of all such models at early times, that is at high
redshifts. Due to the nature of phantom fields, which present an
``inverse'' kinetic behavior in potential slopes, we find that
total energy positivity requires normal power-law potentials
instead of inverse power-law ones (contrary to quintessence case),
with the potential exponent being bounded by the quadratic form.

Furthermore, we have extracted the general cosmological solution
at intermediate times, that is at low redshifts, which is the
period during the transition from matter to dark-energy
domination. Such a solution can be very useful in dark energy
observations, since these (supernovae Ia, WMAP and SDSS ones) are
related to this cosmological period \cite{observ}. The comparison
of the analytic expressions with the exact evolution arising from
numerical elaboration, shows that the tracker solution agrees with
the later within $2\%$ for redshifts $z\gtrsim1.5$, while the
first order solution is accurate within $2\%$ up to $z\approx0.5$.

In the aforementioned analysis we have not used observations in
order to further restrict the potential form, desiring to remain
general. However, from the $w(a)$ expression derived in
(\ref{wtilde}) we conclude that smaller values of $|\alpha|$ are
favored if we want to acquire a $w$ at $z=0.5$ which is not too
negative, consistently with observations within 2$\sigma$. Note
also that smaller $|\alpha|$-values facilitate and enhance the
total energy positivity.

Finally, we have to make a comment about the quantum behavior of
the examined model. As it is known,  the discussion about the
construction of quantum field theory of phantoms is still open in
the literature. For instance in \cite{Cline:2003gs} the authors
reveal the causality and stability problems and the possible
spontaneous breakdown of the vacuum into phantoms and conventional
particles. On the other hand, there have also been serious
attempts in overcoming these difficulties and construct a phantom
theory consistent with the basic requirements of quantum field
theory \cite{quantumphantom0}, with the phantom fields arising as
an effective description. The present analysis is just a first
approach on phantom fields in power-law potentials. Definitely,
the subject of quantization of such models is open and needs
further investigation.\\

\paragraph*{{\bf{Acknowledgements:}}}
The author wishes to thank Institut de Physique Th\'eorique, CEA,
for the hospitality during the preparation of the present work.


\begin{thebibliography}{99}

\bibitem{observ}
A.G. Riess {\it et al.} [Supernova Search Team Collaboration],
Astron. J. {\bf 116}, 1009 (1998);
 S.
Perlmutter {\it{et al.}} [Supernova Cosmology Project
Collaboration], Astrophys. J. {\bf 517}, 565 (1999);
 D. N. Spergel {\it{et al.}}, Astrophys.
J. Suppl. {\bf 148}, 175 (2003); S. W. Allen, {\it{et al.}}, Mon.
Not. Roy. Astron. Soc. {\bf 353}, 457 (2004).

\bibitem{ordishov}
 P. Bin\'{e}truy, C. Deffayet, D. Langlois, Nucl. Phys. B {\bf565}, 269 (2000);
G.R. Dvali, G. Gabadadze, M. Porrati, Phys. Lett. B {\bf485}, 208
(2000); S. Capozziello, Int. J. Mod. Phys. D {\bf11}, 483 (2002);
 S.Nojiri
and S.~D.~Odintsov, Phys. Rev. D {\bf{68}}, 123512 (2003);
  P.~S.~Apostolopoulos, N.~Brouzakis, E.~N.~Saridakis and N.~Tetradis,
  Phys.\ Rev.\  D {\bf 72}, 044013 (2005);
S.Nojiri and S.~D.~Odintsov, Int. J. Geom. Meth. Mod. Phys.
{\bf{4}}, 115 (2007);
  F.~K.~Diakonos and E.~N.~Saridakis,
  arXiv:0708.3143 [hep-th].

\bibitem{quint0}
 P.~J.~E.~Peebles and  B.~Ratra, \apj {\bf 325}, L17 (1988);
 B.~Ratra and P.~J.~E.~Peebles, Phys. Rev. D {\bf 37}, 3406 (1988).

\bibitem{quint}
C.~Wetterich, Nucl.\ Phys.\ B {\bf 302}, 668 (1988); M. S. Turner
and M. White, Phys. Rev. D {\bf{56}}, 4439 (1997); R. R. Caldwell,
R. Dave and P. J. Steinhardt, Phys. Rev. Lett. {\bf{80}}, 1582
(1998); A.~R.~Liddle and R.~J.~Scherrer, Phys.\ Rev.\ D {\bf 59},
023509 (1999); I.~Zlatev, L.~M.~Wang and P.~J.~Steinhardt, Phys.\
Rev.\ Lett.\ {\bf 82}, 896 (1999); Z.~K.~Guo, N.~Ohta and
Y.~Z.~Zhang, Mod.\ Phys.\ Lett.\  A {\bf 22}, 883 (2007);
  O.~Hrycyna and M.~Szydlowski,
  Phys.\ Rev.\  D {\bf 76}, 123510 (2007).

\bibitem{phant} R. R. Caldwell, Phys.
Lett. B {\bf{545}}, 23 (2002); R.~R.~Caldwell, M.~Kamionkowski and
N.~N.~Weinberg, Phys. Rev. Lett. {\bf 91}, 071301 (2003);
  S.~Nojiri and S.~D.~Odintsov,
  Phys.\ Lett.\  B {\bf 562}, 147 (2003);
 V. K.
Onemli and R. P. Woodard, Phys.\ Rev.\ D {\bf 70}, 107301 (2004);
  P.~F.~Gonzalez-Diaz and C.~L.~Siguenza,
  Nucl.\ Phys.\  B {\bf 697}, 363 (2004);
S. Nojiri and S. D. Odintsov, Phys. Rev. D  {\bf{72}}, 023003
(2005);  H. Garcia-Compean, G. Garcia-Jimenez,  O. Obregon, and C.
Ramirez, JCAP 0807, 016 (2008);
  E.~N.~Saridakis,
  arXiv:0811.1333 [hep-th];
  M.~Szydlowski and O.~Hrycyna,
  JCAP {\bf 0901}, 039 (2009);
  X.~m.~Chen, Y.~Gong and E.~N.~Saridakis,
  arXiv:0812.1117 [gr-qc].

\bibitem{quintom}
B.~Feng, X.~L.~Wang and X.~M.~Zhang, Phys.\ Lett.\  B {\bf 607},
35 (2005);
Z. K. Guo, {\it{et al.}}, Phys. Lett. B {\bf 608}, 177 (2005);
M.-Z Li, B. Feng, X.-M Zhang, JCAP,  {\bf0512}, 002 (2005);
  Y.~f.~Cai, H.~Li, Y.~S.~Piao and X.~m.~Zhang,
  Phys.\ Lett.\  B {\bf 646}, 141 (2007);
 W.
Zhao and Y. Zhang, Phys. Rev. D {\bf73}, 123509 (2006);
 B. Feng, M.
Li, Y.-S. Piao and X. Zhang, Phys. Lett. B {\bf 634}, 101 (2006);
  M.~R.~Setare and E.~N.~Saridakis,
  Phys.\ Lett.\  B {\bf 668}, 177 (2008);
  M.~R.~Setare and E.~N.~Saridakis,
  JCAP {\bf 0809}, 026 (2008);
  M.~R.~Setare and E.~N.~Saridakis,
  Phys.\ Lett.\  B {\bf 671}, 331 (2009).

\bibitem{Scherrer:2007pu}
  R.~J.~Scherrer and A.~A.~Sen,
  Phys.\ Rev.\  D {\bf 77}, 083515 (2008);
  R.~J.~Scherrer and A.~A.~Sen,
  Phys.\ Rev.\  D {\bf 78}, 067303 (2008);
  M.~R.~Setare and E.~N.~Saridakis,
  arXiv:0810.4775 [astro-ph].

\bibitem{Kneller:2003xg}
  J.~P.~Kneller and L.~E.~Strigari,
  Phys.\ Rev.\  D {\bf 68}, 083517 (2003);
  L.~R.~W.~Abramo and F.~Finelli,
  Phys.\ Lett.\  B {\bf 575}, 165 (2003);
    X.~Zhang,
  Mod.\ Phys.\ Lett.\  A {\bf 20}, 2575 (2005);
  M.~Yashar, B.~Bozek, A.~Abrahamse, A.~J.~Albrecht and M.~Barnard,
  arXiv:0811.2253 [astro-ph].

\bibitem{Watson:2003kk}
  C.~R.~Watson and R.~J.~Scherrer,
  Phys.\ Rev.\  D {\bf 68}, 123524 (2003).

\bibitem{quint01}
I.~Zlatev, L.~Wang, and P.~J.~Steinhardt, Phys. Rev. Lett. {\bf
82}, 896 (1999);
 A.~R.~Liddle and R.~J.~Scherrer, Phys. Rev. D {\bf 59}, 023509
 (1999);
 P.~J.~Steinhardt, L.~Wang, and I.~Zlatev, Phys. Rev. D {\bf 59}, 123504 (1999).

\bibitem{Binetruy:1998rz}
  P.~Binetruy,
  Phys.\ Rev.\  D {\bf 60}, 063502 (1999);
  A.~Masiero, M.~Pietroni and F.~Rosati,
  Phys.\ Rev.\  D {\bf 61}, 023504 (2000).

\bibitem{Efstathiou}
G. Efstathiou, Mon. Not. R. Astron. Soc. {\bf 310}, 842 (1999).

    \bibitem{Cline:2003gs}
    J.~M.~Cline, S.~Jeon and G.~D.~Moore,
    Phys.\ Rev.\  D {\bf 70}, 043543 (2004).

\bibitem{quantumphantom0}
  S.~Nojiri and S.~D.~Odintsov,
  Phys.\ Lett.\  B {\bf 562}, 147 (2003);
  S.~Nojiri and S.~D.~Odintsov,
  Phys.\ Lett.\  B {\bf 571}, 1 (2003).



\end{thebibliography}
\end{document}